\documentclass[prl,twocolumn,superscriptaddress,showpacs,
amsmath,amssymb]{revtex4}

\usepackage{graphicx} 
\usepackage{dcolumn} 

\usepackage[english]{babel}
\usepackage[babel,kerning=true,spacing=true]{microtype}

\def\bk{{\bf k}}

\def\bp{{\bf p}}

\def\bQ{{\bf Q}}

\def\b0{{\bf 0}}

\def\bra{\langle}
\def\ket{\rangle}
\def\up{\uparrow}
\def\down{\downarrow}

\def\eps{\epsilon}

\def\Gam{\Gamma}

\def\Lam{\Lambda}

\def\sg{\sigma}

\def\sgn{{\rm sgn}}

\begin{document}

\title{Coexistence of incommensurate magnetism and superconductivity in the two-dimensional Hubbard model}

\author{Hiroyuki Yamase}
\affiliation{Max Planck Institute for Solid State Research,
 D-70569 Stuttgart, Germany}
\affiliation{National Institute for Materials Science, Tsukuba 305-0047, Japan}
\author{Andreas Eberlein}
\affiliation{Max Planck Institute for Solid State Research,
 D-70569 Stuttgart, Germany}
\affiliation{Department of Physics, Harvard University, Cambridge MA 02138, USA}
\author{Walter Metzner}
\affiliation{Max Planck Institute for Solid State Research,
 D-70569 Stuttgart, Germany}

\date{\today}

\begin{abstract}
We analyze the competition of magnetism and superconductivity in the two-dimensional Hubbard model with a moderate interaction strength, including the possibility of incommensurate spiral magnetic order.
Using an unbiased renormalization group approach, we compute magnetic and superconducting order parameters in the ground state.
In addition to previously established regions of N\'eel order coexisting with $d$-wave superconductivity, the calculations reveal further coexistence regions where superconductivity is accompanied by incommensurate magnetic order.
\end{abstract}

\maketitle


{\em Introduction.}
The two-dimensional Hubbard model is a prototype system for competing order in layered transition metal oxide compounds. Shortly after the discovery of high temperature superconductivity in cuprates, it has been proposed as a key model describing the valence electrons in the copper-oxygen planes \cite{anderson87}.
Indeed, the model exhibits the most prominent ordering phenomena observed in high-$T_c$ cuprates, namely antiferromagnetism and $d$-wave superconductivity \cite{scalapino12}.

While the magnetic order is captured already by conventional mean-field theory \cite{montorsi92}, superconductivity is fluctuation-driven and hence more subtle.
Nevertheless, the emergence of $d$-wave superconductivity in the 2D Hubbard model is nowadays well established \cite{scalapino12}. In particular, unbiased evidence for superconductivity with sizable gaps at moderate interaction strengths has been obtained from functional renormalization group (fRG) calculations \cite{metzner12,friederich11,eberlein14}, and from embedded quantum cluster methods at intermediate and strong coupling \cite{maier00,lichtenstein00,capone06,aichhorn06,khatami08,gull13,zheng15}.

The magnetic order is not necessarily of commensurate N\'eel type, that is, with antiparallel spin orientation between adjacent sites.
The possibility of magnetic order with generally {\em incommensurate}\/ wave vectors away from $(\pi,\pi)$ has been explored by several mean-field studies \cite{machida89,schulz90,dombre90,fresard91,igoshev10}, and also by expansions in the limit of a small hole density, where fluctuation effects were taken into account \cite{shraiman89,chubukov,sushkov04}.
Incommensurate magnetic order in the two-dimensional Hubbard model is also indicated by diverging interactions and susceptibilities at incommensurate momenta in fRG flows \cite{metzner12,halboth00,husemann09}.
However, the competition and possible coexistence of incommensurate magnetism and superconductivity has not yet been analyzed \cite{footnote1}.
To do this, one needs to access the ordered phase in a framework that captures the fluctuations which generate $d$-wave superconductivity, allowing at the same time for a high momentum space resolution to distinguish the incommensurate ordering wave vector from $(\pi,\pi)$.
The latter requirement is an obstacle for cluster methods, which have so far been restricted to commensurate antiferromagnetism \cite{lichtenstein00,capone06,aichhorn06,zheng15}.

In the fRG flow spontaneous symmetry breaking is signaled by diverging effective interactions at a critical energy scale $\Lam_c$.
In principle, the flow can be continued below the instability scale $\Lam_c$ to compute order parameters in the ordered phase \cite{metzner12,salmhofer04,baier04}.
However, this is rather complicated, especially in case of two or more order parameters.
A drastic simplification occurs if the flow below the scale $\Lam_c$ is approximated by mean-field theory (MFT), since then order parameters can be computed without dealing with anomalous interactions \cite{reiss07}.
While fluctuations above $\Lam_c$ are crucial for the generation of $d$-wave superconductivity, fluctuations below $\Lam_c$ are not expected to affect ground state order parameters significantly.
Recently, a consistent fusion of the fRG flow above the scale $\Lam_c$ with MFT below $\Lam_c$ has been formulated and used to study the competition of antiferromagnetism and superconductivity in the Hubbard model \cite{wang14}.
In the purely superconducting regime of the ground state phase diagram, the pairing gap computed by fRG+MFT could be compared to previous results from a complete fRG flow with symmetry breaking (including fluctuations below $\Lam_c$) \cite{eberlein14}. The results obtained from the two methods agree very well.
Away from half-filling, antiferromagnetic order was shown to be accompanied by microscopically coexisting $d$-wave superconductivity, in qualitative agreement with results from a previous fRG+MFT computation \cite{reiss07} and from cluster methods at stronger interactions \cite{lichtenstein00,capone06,aichhorn06,zheng15}.

Here we present novel results for the two-dimensional Hubbard model, allowing, for the first time, magnetic order with arbitrary wave vectors and superconductivity. In addition to the previously established regions of N\'eel order coexisting with $d$-wave superconductivity, we find new coexistence regions where superconductivity is accompanied by incommensurate magnetic order.


{\em Method.}
Our results are based on the fRG+MFT method formulated in Ref.~\cite{wang14}.
The calculation consists of three steps.
First, the two-particle vertex is computed from a fRG flow integrated down to a scale $\Lam_{\rm MF}$ slightly above the critical scale $\Lam_c$.
The flow is approximated by a one-loop truncation, with a static (frequency independent) vertex parametrized via a decomposition in charge, magnetic, and pairing channels \cite{husemann09} with $s$-wave and $d$-wave form factors as described in Ref.~\cite{eberlein14}.
The scale dependence defining the fRG flow is introduced by a soft frequency cutoff regularizing the bare propagator as
\begin{equation}
G_0^{\Lam}(k_0,\bk) =
 \Big[i \, \sgn(k_0) \sqrt{k_0^2 + \Lam^2} - (\eps_{\bk} - \mu) \Big]^{-1} \, .
\end{equation}
We denote the spin and momentum dependence of the vertex at scale $\Lam$ by
$\Gam^{\Lam}_{\sg_1,\sg_2;\sg_3,\sg_4}(\bk_1,\bk_2;\bk_3,\bk_4)$, where the indices 1,2 refer to outgoing, and 3,4 to incoming particles.
Spin-singlet pairing is driven by the singlet component of the vertex in the Cooper channel (zero total momentum),
\begin{equation} \label{V}
 V_{\bk\bk'} = 
 \frac{1}{2} \Gam^{\Lam_{\rm MF}}_s(\bk,-\bk;-\bk',\bk') \, ,
\end{equation}
where $\Gam^{\Lam}_s = \Gam^{\Lam}_{\sg,-\sg;-\sg,\sg} - \Gam^{\Lam}_{\sg,-\sg;\sg,-\sg}$.
Magnetic instabilities with a momentum $\bQ$ are generated by the magnetic component of the vertex with momentum transfer $\bQ$, that is,
\begin{equation} \label{U}
 U_{\bQ;\bk\bk'} =
 \Gam^{\Lam_{\rm MF}}_{\sg,-\sg;\sg,-\sg}(\bk+\bQ,\bk';\bk'+\bQ,\bk) \, .
\end{equation}
Note that
$\Gam^{\Lam}_{\sg,-\sg;\sg,-\sg} =
 \Gam^{\Lam}_{\sg,\sg;\sg,\sg} - \Gam^{\Lam}_{\sg,-\sg;-\sg,\sg}$
due to spin rotation invariance.

In the second step, the {\em irreducible}\/ vertices in the relevant channels are computed by solving the corresponding Bethe-Salpeter equations \cite{wang14,supp1}.
The irreducible pairing vertex $\tilde V_{\bk\bk'}$ is obtained from the equation
\begin{eqnarray} \label{V_irr}
 V_{\bk\bk'} &=& \tilde V_{\bk\bk'}
 \nonumber \\
 &-& \int_p
 \tilde V_{\bk\bp}
 G^{\Lam_{\rm MF}}(p_0,\bp) G^{\Lam_{\rm MF}}(-p_0,-\bp) 
 V_{\bp\bk'} \, ,
\end{eqnarray}
where $G^{\Lam_{\rm MF}}(p_0,\bp)$ is the propagator at the scale
$\Lam_{\rm MF}$, and $\int_p$ is a short-hand notation for
$\int \frac{d^2\bp}{(2\pi)^2} \frac{dp_0}{2\pi}$.
The irreducible magnetic vertex $\tilde U_{\bQ;\bk\bk'}$ is determined from
\begin{eqnarray} \label{U_irr}
 U_{\bQ;\bk\bk'} &=& 
 \tilde U_{\bQ;\bk\bk'}
 \nonumber \\
 &+& \int_p \tilde U_{\bQ;\bk\bp}
 G^{\Lam_{\rm MF}}(p_0,\bp) G^{\Lam_{\rm MF}}(p_0,\bp\!+\!\bQ)
 U_{\bQ;\bp\bk'} \, . \nonumber \\
\end{eqnarray}
In this work we neglect self-energy contributions to the flow, so that $G^{\Lam_{\rm MF}}$ is just the regularized bare propagator $G_0^{\Lam_{\rm MF}}$.

Finally, the superconducting and magnetic order parameters are computed by solving the mean-field Hamiltonian with the irreducible vertex parts $\tilde V_{\bk\bk'}$ and $\tilde U_{\bQ;\bk\bk'}$ as effective interactions.
The superconducting order parameter is given by the gap function
\begin{equation}
 \Delta_{\bk} = \int \! \frac{d^2\bk'}{(2\pi)^2} \,
 \tilde V_{\bk\bk'} \bra p_{\bk'} \ket \, ,
\end{equation}
where $p_{\bk} = a_{-\bk\down} a_{\bk\up}$ is the Cooper pair annihilation operator.
The phase of the superconducting order can be chosen such that $\Delta_{\bk}$ is real.
Concerning the magnetic order, we restrict our analysis to spiral states, as described by the order parameter
\begin{equation}
 A_{\bk} = \int \! \frac{d^2\bk'}{(2\pi)^2} \,
 \tilde U_{\bQ;\bk\bk'} \bra m_{\bk'} \ket \, ,
\end{equation}
where $m_{\bk} = a_{\bk\up}^{\dag} a_{\bk+\bQ\down}$.
For $\bQ = (\pi,\pi)$ the spiral order is equivalent to N\'eel order with a staggered magnetisation in the $xy$-plane.
There are several mean-field studies of spiral magnetic order in the two-dimensional Hubbard model \cite{fresard91,igoshev10}.
A frequently discussed alternative is collinear order \cite{machida89,schulz90}, especially in combination with pronounced charge stripes \cite{zaanen89,kato90,fleck01}.
In a spiral state the amplitude of the magnetization is homogeneous, only the orientation varies. Hence, the magnetic order entails an energy gain everywhere in the system.
By contrast, collinear order necessarily involves regions with a reduced magnetization, where the energy gain from the order is also reduced. Incommensurate collinear order is therefore expected to be favorable only in the form of sharply defined charged domain walls between antiferromagnetic regions \cite{machida89,schulz90}.
In the moderate interaction regime investigated here such pronounced real space profiles seem unfavorable since they cost a lot of kinetic (hopping) energy.

A mean-field decoupling of the reduced effective interactions yields the mean-field Hamiltonian
\begin{eqnarray}
 H_{\rm MF} &=& \sum_{\bk,\sg} \eps_{\bk} a_{\bk\sg}^{\dag} a_{\bk\sg}
 \nonumber \\
 &+& \sum_{\bk} A_{\bk} m_{\bk}^{\dag} + A_{\bk}^* m_{\bk}
 - A_{\bk}^* \bra m_{\bk} \ket \nonumber \\
 &+& \sum_{\bk} \Delta_{\bk} p_{\bk}^{\dag} + \Delta_{\bk}^* p_{\bk}
 - \Delta_{\bk}^* \bra p_{\bk} \ket \, .
\end{eqnarray}
For the Hubbard model with nearest and next-to-nearest neighbor hopping on a square lattice, the dispersion relation is
$\eps_{\bk} = -2t (\cos k_x + \cos k_y) - 4t' \cos k_x \cos k_y$.
Using Nambu spinors $\Psi_{\bk} = (a_{\bk\up}^{\phantom\dag}, a_{-\bk\down}^{\dag}, a_{\bk+\bQ\down}^{\phantom\dag}, a_{-\bk-\bQ\up}^{\dag})$, the mean-field Hamiltonian can be written in the form
$H_{\rm MF} = \frac{1}{2} \sum_{\bk} \Psi_{\bk}^{\dag} {\cal M}_{\bk} \Psi_{\bk} + {\rm const}$, where ${\cal M}_{\bk}$ is a hermitian $4 \times 4$ matrix.
$H_{\rm MF}$ can thus be diagonalized by a $4 \times 4$ unitary (generalized Bogoliubov) transformation, and the resulting gap equations can be solved numerically by iteration.
Occasionally two distinct locally stable solutions of the gap equations are found. One then has to compute the corresponding free energies to discriminate globally stable from metastable states.


{\em Results.}
We have computed the magnetic and superconducting order parameters in the ground state of the two-dimensional Hubbard model at and near half-filling for weak to moderate interaction strenghts $U/t = 2,3,4$ for a small next-to-nearest neighbor hopping amplitude $t'/t=-0.15$, and for $U/t=3$ also for the special particle-hole symmetric case $t'=0$.
We now present our results for $U/t=3$ and $t'/t=-0.15$, and subsequently discuss similarities and differences found for the other choices of $U$ and $t'$ \cite{supp2}.
In the following we set $t=1$.

In Fig.~1 we show the amplitudes of the magnetic and superconducting order parameters, $A = \max_{\bk} A_{\bk}$ and $\Delta = \max_{\bk} \Delta_{\bk}$, as a function of the electron density.
\begin{figure}
\begin{center}
\includegraphics[width=7cm]{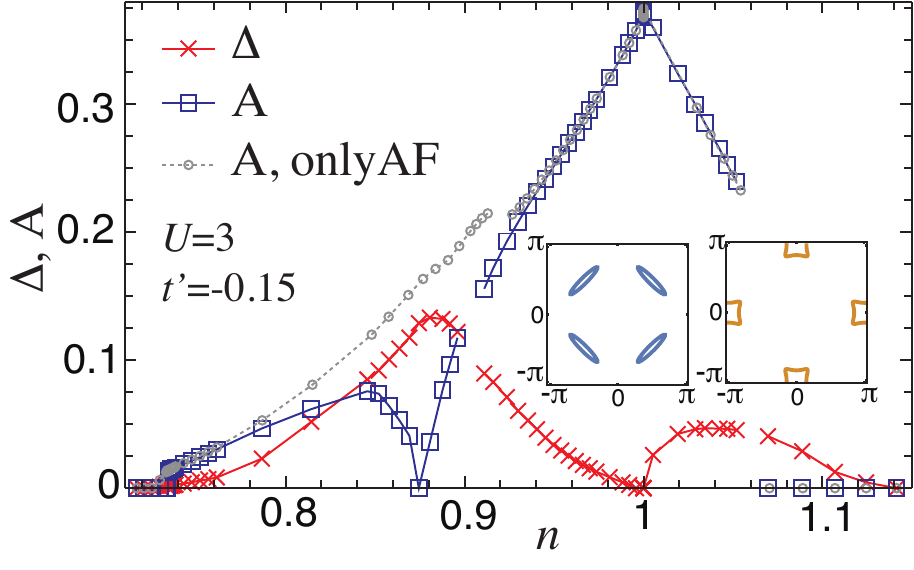}
\caption{(Color online) Amplitudes of magnetic and superconducting
 gap functions in the ground state of the two-dimensional Hubbard
 model as a function of the electron density $n$, for $U=3$ and
 $t'=-0.15$. The amplitude of the magnetic gap function obtained
 from a purely magnetic solution is also shown for comparison, with
 corresponding hole and electron pockets at $n=0.96$ and $n=1.05$
 in the left and right insets, respectively.
 First order transitions (see broken lines near $n=0.9$ and $n=1.06$)
 lead to small density intervals where no homogeneous solution
 exists.}
\end{center}
\end{figure}
The magnetic order is of N\'eel-type ($\bQ = (\pi,\pi)$) at half-filling. The N\'eel order persists on the electron-doped side ($n>1$), and also for small and moderate hole doping ($0.9<n<1$).
The stability of commensurate antiferromagnetic order on the electron-doped side is generally expected \cite{rowe12}.
For $n<1$ excitations in hole-pockets near the Brillouin zone diagonal could destabilize the commensurate state even for small hole doping for small $|t'|$ and/or large $U$ \cite{chubukov}.
At $n = 0.9$ a first order transition to an incommensurate spiral state occurs, with a wave vector of the form $\bQ = (\pi - 2\pi\eta,\pi)$, or equivalent (by symmetry) wave vectors $(\pi + 2\pi\eta,\pi)$ or $(\pi,\pi \pm 2\pi\eta)$.
The magnetic order is completely suppressed by superconductivity at van Hove filling ($n=0.87$), but then reemerges for lower densities.
The magnetic transition at $n=0.73$ is of weak first-order type.
The incommensurability $\eta$ is plotted as a function of the density in Fig.~2.
It jumps from zero to a small finite value at the commensurate-incommensurate transition, and then increases monotonically upon further doping until the magnetic order disappears at $n = 0.73$.
The chemical potential $\mu(n)$, which is also plotted in Fig.~2, exhibits a discontinuity due to the charge gap at half-filling. In other words, the density $n$ is pinned to half-filling for a range of chemical potentials between $-0.37$ and $-0.245$.
\begin{figure}
\begin{center}
\includegraphics[width=7.8cm]{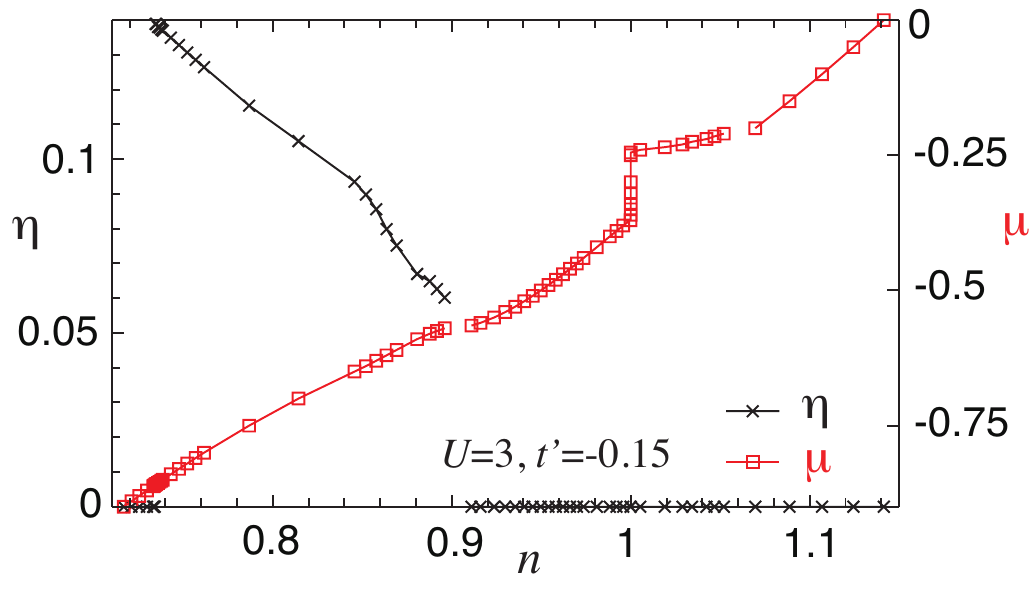}
\caption{(Color online) Incommensurability $\eta$ and chemical potential $\mu$
 as function of the density for the same parameters as in Fig.~1.}
\end{center}
\end{figure}
Comparing to the purely magnetic solution (excluding superconductivity), which is also shown in Fig.~1, one can see that superconductivity has little influence on the magnetic order in the regime around half-filling where it is commensurate. 
The incommensurate magnetic order on the hole-doped side is strongly suppressed by pairing in the vicinity of van Hove filling. At van Hove filling, superconductivity eliminates the magnetic order completely.
By contrast, in the overdoped regime well below van Hove filling, incommensurate magnetic order and superconductivity coexist without suppressing each other significantly.

The pairing gap $\Delta_{\bk}$ is finite for all densities except at half-filling, where the Fermi surface is fully gapped by the antiferromagnetic order.
Hence, away from half-filling, magnetic order always allows for coexisting superconductivity. This is easily understood as follows.
Doping the half-filled antiferromagnet by additional electrons or holes leads to electron or hole pockets (see the insets of Fig.~1). The ubiquitous attractive $d$-wave pairing interaction inevitably generates a Cooper instability at the (small) Fermi surfaces enclosing these pockets, and thus superconductivity.
The hole pockets for $n<1$ are centered around the nodes of the pairing gap $\Delta_{\bk}$, while the electron pockets for $n>1$ are in the antinodal region near $(\pi,0)$ and $(0,\pi)$, where the gap is maximal. Hence, the onset of pairing around half-filling is much steeper on the electron doped side (see Fig.~1).
This is in agreement with recent spin-fluctuation exchange calculations in the weak coupling regime \cite{rowe15}, but differs from the behavior found in a strong coupling analysis of electrons moving in an antiferromagnetic spin background.
In the latter case, pairing is mediated mostly by {\em transverse}\/ spin fluctuations (magnons), which couple very weakly to electrons in the antinodal region, so that the pairing interaction is very small for low electron doping \cite{kuchiev95}. By contrast, at moderate coupling also {\em longitudinal} spin fluctuations contribute, and yield a sizable pairing interaction in the antinodal region \cite{rowe15}.

The maximal gap amplitude at ''optimal'' hole-doping is significantly larger than the maximal gap on the electron doped side, in agreement with the gap hierarchy in cuprate superconductors.
The maximal gap on the hole-doped side is situated slightly above van Hove filling, where the magnetic order is already quite weak.
The gap decreases smoothly in the ''overdoped'' regime, due to a decrease of the pairing interaction and the density of states.

The leading order parameter is often guessed from the leading divergence of the effective interactions in the fRG flow upon approaching the critical scale $\Lam_c$ \cite{metzner12}. This is usually correct, but there are exceptions. In particular, around van Hove filling, we find superconductivity as the dominant order, although the leading divergence occurs actually in the magnetic channel (see Ref.~\cite{eberlein14} where the fRG flow was studied for the same parameters).

A major novel result of our work is the coexistence of superconductivity with incommensurate magnetic order in a regime where a purely superconducting state was expected, since the hole-doping is already quite large.
It is therefore interesting to look at the condensation energy gained by symmetry breaking in the incommensurate regime.
In Fig.~3 we plot the total condensation energy $E(A,\Delta) - E(0,0)$ gained by magnetism and pairing.
\begin{figure}
\begin{center}
\includegraphics[width=7cm]{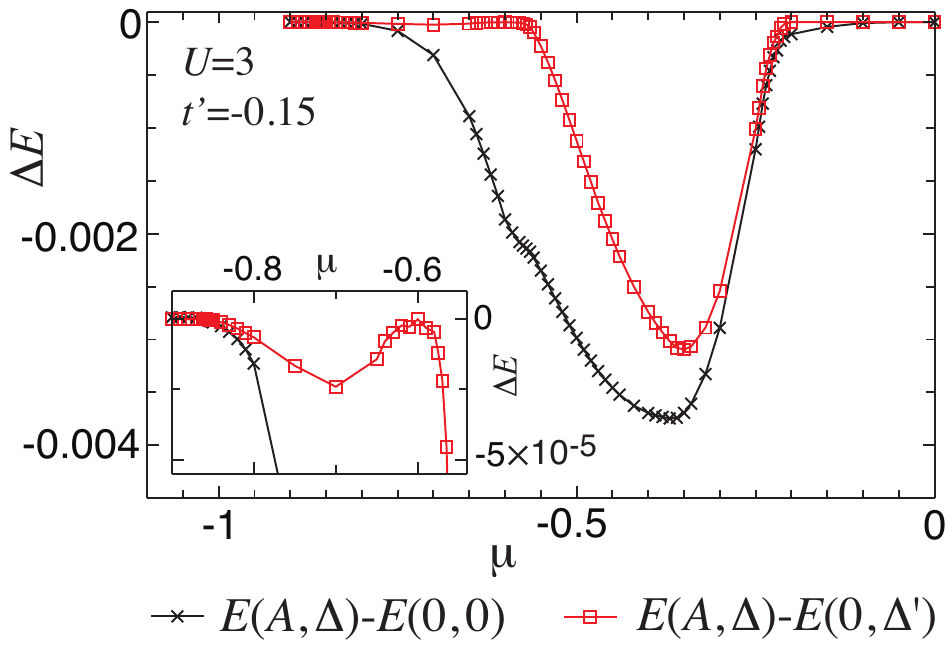}
\caption{(Color online) Total condensation energy $E(A,\Delta) - E(0,0)$ and relative magnetic condensation energy $E(A,\Delta) - E(0,\Delta')$ as a function of the chemical potential $\mu$. The inset is a zoom into the region with incommsurate magnetic order. The model parameters are the same as in Fig.~1.}
\end{center}
\end{figure}
We also plot the additional energy gain due to the magnetic order compared to a purely superconducting state, $E(A,\Delta) - E(0,\Delta')$, where $\Delta'$ is the pairing gap in the absence of magnetism.
One can see that this additional energy gain is tiny in the incommensurate regime ($\mu < -0.57$), even for densities where the size of the magnetic order parameter is comparable to the pairing gap.
Hence, the magnetic order in this regime is extremely delicate, a kind of {\em "gossamer magnetism"}, reminiscent of Laughlin's "gossamer superconductivity" in lightly doped Mott insulators \cite{laughlin02}.
The quasi-degeneracy of a purely superconducting state and a state with coexisting magnetism can be expected to lead to intriguing fluctuation effects.

Let us now compare the above results to those obtained for other choices of $U$, as presented in our Supplementary Material \cite{supp2}.
For $U=2$ and $t'=-0.15$, the ground state is purely superconducting for all densities, with a small $d$-wave gap which is maximal slightly above van Hove filling. Magnetic order occurs only if superconductivity is switched off. This is expected in the weak coupling limit in the absence of nesting (for finite $t'$).
For $U=4$ and $t'=-0.15$ there is no homogeneous solution in the density range $0.88 < n < 1$, which includes also van Hove filling. Hence, a system with an average density in that interval will undergo phase separation in regions with distinct densities $n=1$ and $n=0.88$, or form a more complex type of order. For smaller densities incommensurate magnetic order coexisting with $d$-wave superconductivity is found, again with a tiny energy gain from the magnetic order.
Hence, our main result, the coexistence of superconductivity with a very delicate incommensurate magnetic order at sizable hole-doping, is robust with respect to an increase of $U$. On the electron-doped side, there are no qualitative differences compared to $U = 3$, except for a tiny incommensurate region at the edge of the magnetic regime.

For $t'=0$, the Hubbard model is particle-hole symmetric, that is, the properties for densities $n$ and $2-n$ are equivalent. Due to perfect nesting the system is a N\'eel antiferromagnet at half-filling for any $U > 0$. There is no homogeneous solution in a density range around half-filling, both on the electron and hole-doped side.
For larger doping there is a small region exhibiting incommensurate magnetism in coexistence with $d$-wave superconductivity, and a broader purely superconducting region.
While the magnetic order parameter at half-filling has almost the same size for $t'=0$ and $t' = -0.15$, the largest achievable pairing gap (at optimal doping) is much smaller for $t'=0$. A sizable next-to-nearest neighbor hopping thus helps to promote superconductivity with a large gap.
This was already revealed in a previous fRG study \cite{eberlein14} and in a recent quantum cluster calculation \cite{zheng15}, and is in qualitative agreement with the empirical trend in cuprates \cite{pavarini01}.


{\em Summary.}
We have analyzed the competition between magnetism and superconductivity in the ground state of the two-dimensional Hubbard model, including the possibility of incommensurate spiral magnetic order.
Using a combination of fRG and mean-field theory, fluctuation driven order is captured without any bias for a specific instability. Charge, spin, and pairing channels are treated on equal footing.
Away from half-filling, magnetic order always coexists with superconductivity, as a consequence of a Cooper instability in electron or hole pockets.
For $t' < 0$, both magnetism and superconductivity exhibit a pronounced particle-hole asymmetry.
On the hole-doped side, superconductivity has larger maximal gaps and it coexists with incommensurate magnetism at moderate doping, except at van Hove filling, where magnetic order is fully suppressed by pairing.
The incommensurate magnetic order is ``gossamer-like'' in the sense that it is stabilized only by a tiny energy gain with respect to a purely superconducting state.
Rather fragile incommensurate magnetic order has actually been observed at the bottom of the superconducting dome in $\rm La_{2-x} Sr_x Cu O_4$ \cite{wakimoto01} and $\rm Y Ba_2 Cu_3 O_{6+x}$ \cite{miller06,haug10}.
Suppressing superconductivity by a strong magnetic field would stabilize that order.
Indeed, in a very recent high field experiment a Hall coefficient consistent with hole-pockets arising from possible spin-density waves was observed in $\rm Y Ba_2 Cu_3 O_{6+x}$ at much higher doping than previously \cite{badoux15}.

Our analysis was restricted to weak and moderate interaction strengths. The method may be extended to strong interactions by using dynamical mean-field theory as a starting point for the fRG flow \cite{taranto14}.


\vskip 2mm

\begin{acknowledgments}
We would like to thank A.~Chubukov, A.~Katanin, O.~Sushkov, and R.~Zeyher for valuable discussions.
Support from the DFG research group FOR 723 is gratefully acknowledged.
HY also appreciates support by the Alexander von Humboldt Foundation and a Grant-in-Aid for Scientific Research from the Japan Society for the Promotion of Science.
AE acknowledges partial support by the German National Academy of Sciences Leopoldina through grant LPDS~2014-13
\end{acknowledgments}



\setcounter{equation}{0}
\setcounter{figure}{0}
\setcounter{table}{0}
\makeatletter
\renewcommand{\theequation}{S\arabic{equation}}
\renewcommand{\thefigure}{S\arabic{figure}}
\renewcommand{\thetable}{S\arabic{table}}
\renewcommand{\bibnumfmt}[1]{[S#1]}
\renewcommand{\thesection}{\Alph{section}}

\onecolumngrid

\newpage


\setcounter{page}{1}

\section{Coexistence of incommensurate magnetism and superconductivity in the
 two-dimensional Hubbard model: Supplemental material}
\begin{center}
\vspace*{-6pt}
H.~Yamase, A.~Eberlein, and W. Metzner
\end{center}

\vskip 5mm


In this Supplemental Material we provide details on the computation of the irreducible vertices, and we present additional results for alternative choices
of the model parameters.

\subsection{\bf Computation of irreducible vertices}

Here we discuss some technical aspects concerning the computation of the irreducible vertices $\tilde V_{\bk\bk'}$ and $\tilde U_{\bQ;\bk\bk'}$ from $V_{\bk\bk'}$ and $U_{\bQ;\bk\bk'}$, respectively.
The $\bk$ and $\bk'$ dependences are discretized by partitioning the momentum space in patches, that is,
\begin{equation}
 \begin{array}{rcl}
 V_{\bk\bk'} &=& 
 \sum_{j,j'} V_{jj'} \Theta_j(\bk) \Theta_{j'}(\bk') \, , \\
 U_{\bQ;\bk\bk'} &=&
 \sum_{j,j'} U_{\bQ;jj'} \Theta_j(\bk) \Theta_{j'}(\bk') \, ,
 \end{array}
\end{equation}
where $\Theta_j(\bk)$ is the characteristic function of the patch labelled by $j$, that is, $\Theta_j(\bk) = 1$ if $\bk$ is on patch $j$ and $\Theta_j(\bk) = 0$ otherwise.
The solution of the Bethe-Salpether equations is thus reduced to a matrix inversion.
The matrices $V_{jj'}$ and $U_{\bQ;jj'}$ are real and symmetric in $j$ and $j'$. Hence, they can be decomposed into eigenmodes as
$V_{jj'} = \sum_n v^{(n)} e^{(n)}_j e^{(n)}_{j'}$ and
$U_{\bQ;jj'} = \sum_n u^{(n)}_{\bQ} e^{(n)}_{\bQ;j} e^{(n)}_{\bQ;j'}$
with real eigenvalues and real normalized eigenvectors.
The leading instabilities correspond to the most negative eigenvalues.
However, the discretized irreducible vertices $\tilde V_{jj'}$ and $\tilde U_{\bQ;jj'}$ obtained from the Bethe-Salpeter equations contain eigenmodes with large negative eigenvalues which correspond to eigenmodes with large {\em positive} eigenvalues of the full vertices.
To remove these unphysical instabilities, we project the vertices on the subspace with negative eigenvalues,
\begin{equation}
 \begin{array}{rcl} 
 V'_{jj'} &=& \sum'_n v^{(n)} e^{(n)}_j e^{(n)}_{j'} \, , \\
 U'_{\bQ;jj'} &=& \sum'_n u^{(n)}_{\bQ} e^{(n)}_{\bQ;j} e^{(n)}_{\bQ;j'} \, ,
 \end{array}
\end{equation}
where the primed sums are restricted to {\em negative}\/ eigenvalues.
The irreducible vertices $\tilde V'_{jj'}$ and $\tilde U'_{\bQ;jj'}$ computed from the projected full vertices do not contain any unphysical instabilities, and the leading physical instabilities are not affected by the projection.


\subsection{\bf Results for alternative choices of model parameters}

In this section we present results for alternative choices of the model parameters, which complement the results for $t'/t = -0.15$ and $U/t = 3$ presented in the main paper. In the following we set $t=1$.

In Fig.~S1 we show the amplitudes of magnetic and superconducting order parameters,
$A = \max_{\bk} A_{\bk}$ and $\Delta = \max_{\bk} \Delta_{\bk}$, as a function of the
electron density for $t'=-0.15$ and two distinct interaction strengths, $U=2$ (left) and $U=4$ (right).
\begin{figure}[htb]
\begin{center}
\includegraphics[width=7.2cm]{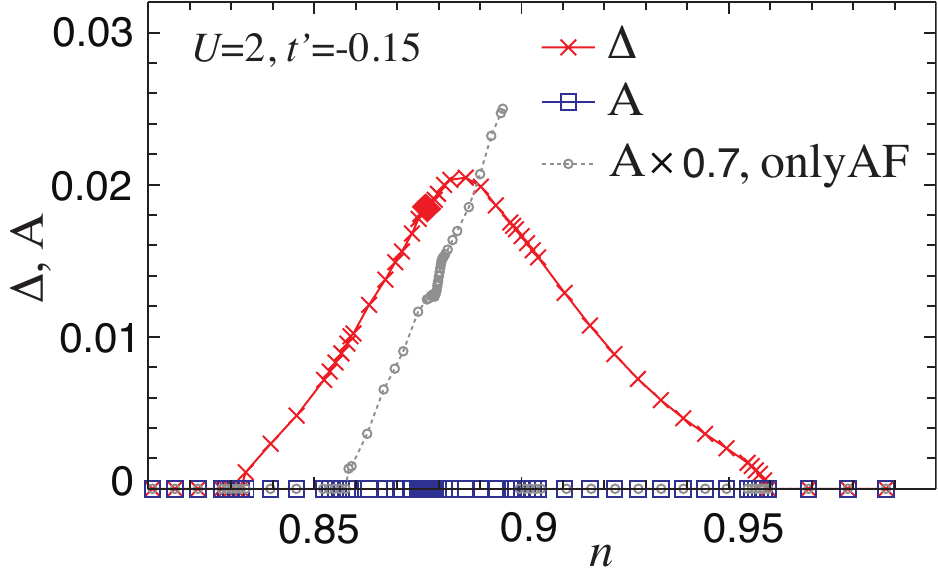} \hskip 1cm
\includegraphics[width=7cm]{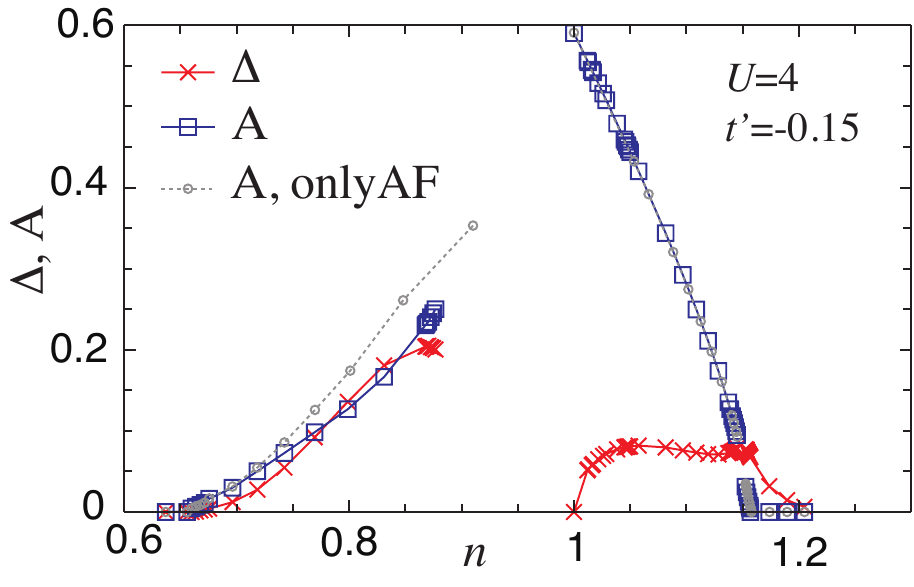}
\caption{Amplitudes of magnetic and superconducting gap
 functions in the ground state of the two-dimensional Hubbard model for
 a next-to-nearest neighbor hopping amplitude $t'=-0.15$ and interaction
 strengths $U=2$ (left) and $U=4$ (right).
 The amplitude of the magnetic gap function obtained from a purely
 magnetic solution is also shown for comparison.}
\end{center}
\end{figure}
For $U=2$, the minimal energy is reached by a purely superconducting solution at all densities. The $d$-wave gap function has a sizable amplitude only in a density range between $n=0.83$ and $n=0.96$, with a maximum slightly above van Hove filling ($n_{\rm vH} = 0.875$).
In the absence of superconductivity (setting $\Delta_{\bk} = 0$ in the mean-field equations), a magnetic solution emerges in a small density range around van Hove filling. The magnetic order is commensurate for densities $n > 0.879$ and incommensurate below. The commensurate-incommensurate transition is thus situated slightly above van Hove filling. It is discontinuous, with a very small jump of the magnetic order parameter (not visible in the figure).
The deviation of the ordering wave vector from $(\pi,\pi)$ is quite small ($\eta < 0.04$).
The incommensurate magnetic transition near $n=0.86$ is continuous, the commensurate transition at $n=0.895$ is discontinuous.

For $U=4$ the model exhibits N\'eel order with a sizable magnetic order parameter at half-filling. The Fermi surface is fully gapped, and the corresponding charge gap leads to a pinning of the density at half-filling for a range of chemical potentials from $\mu=-0.64$ to $\mu=-0.036$. While the ground state should be invariant in that range, our approximate results for the magnetic order parameter $A_{\bk}$ depend to some extent on the choice of $\mu$. In Fig.~S1 we have plotted $A$ as obtained at $\mu = -0.036$. At the other edge, $A$ is about 10 percent larger. This rather strong increase occurs near the van Hove singularity of the non-interacting system at $\mu =-0.6$. For $U=3$ the artificial dependence on $\mu$ is much smaller (below one percent).
At $\mu=-0.64$ there is a first order transition between the N\'eel state and a phase with incommensurate magnetic order coexisting with $d$-wave superconductivity. The transition is accompanied by a large density drop to $n=0.88$. No homogeneous solution exists for densities between $n=0.88$ and $n=1$, so that systems with densities in that range must undergo phase separation.
The magnetic order parameter vanishes continuously at $n=0.66$.
In the entire density range $0.66 < n < 0.88$ the incommensurate magnetic order coexists with $d$-wave superconductivity. While the amplitudes of the two order parameters are generally comparable in this coexistence region, the energy gain resulting from the magnetic order is tiny compared to the gain obtained by pairing (see the inset of Fig.~S2). This ``gossamer-type'' magnetism was already observed for $U=3$ (see Fig.~3).
\begin{figure}[htb]
\begin{center}
\includegraphics[width=7cm]{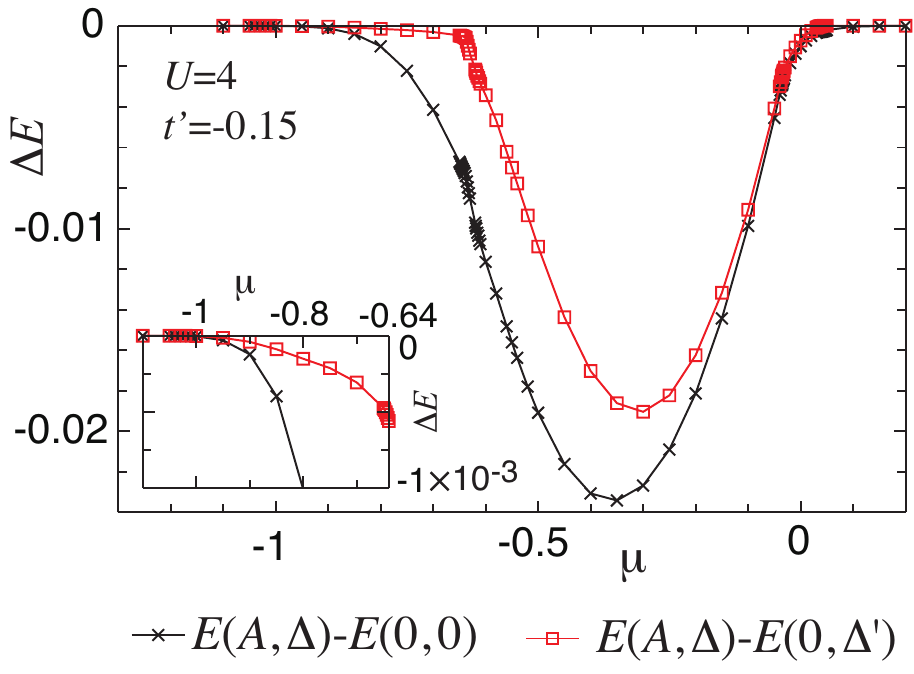}
\caption{Total condensation energy $E(A,\Delta) - E(0,0)$ and relative
 magnetic condensation energy $E(A,\Delta) - E(0,\Delta')$ as a function
 of the chemical potential $\mu$, for $t'=-0.15$ and $U=4$.}
\end{center}
\end{figure}
On the electron-doped side the commensurate magnetic order persists in a wide density range. There is a first order transition to an incommensurate state with a tiny incommensurability at $n=1.15$, near the edge of the magnetic region. The magnetic order parameter vanishes continuously at a slightly higher density.
The magnetic order on the electron-doped side also coexists with $d$-wave superconductivity, with a steep onset of pairing near half-filling. The pairing gap remains sizable in a small density range outside the magnetic regime.

We now turn to the special case with pure nearest-neighbor hopping ($t'=0$) for comparison. The Fermi surface at half-filling is perfectly nested in that case. Hence, at half-filling the system is a N\'eel antiferromagnet for any $U > 0$.
The system is particle-hole symmetric, that is, the properties for densities $n$ and $2-n$ are equivalent, so that we can restrict ourselves to the hole-doped region.
We have computed the magnetic and superconducting order parameters for $U = 3$.
At half-filling the system is purely antiferromagnetic with a magnetic order parameter $A = 0.36$.
No homogeneous solution is found in the density range $0.875 < n < 1$. Hence, a system with an average density in that interval will undergo phase separation in regions with distinct densities $n=1$ and $n=0.875$. 
The magnetic and superconducting order parameters for $n < 0.875$ are shown in Fig.~S3.
For $n < 0.875$ there is a small region exhibiting incommensurate magnetism in coexistence with $d$-wave superconductivity, and a broader purely superconducting region below. The incommensurability is small ($\eta = 0.041$) and almost density-independent in the narrow regime where the incommensurate magnetic order exists.
\begin{figure}[htb]
\begin{center}
\includegraphics[width=7cm]{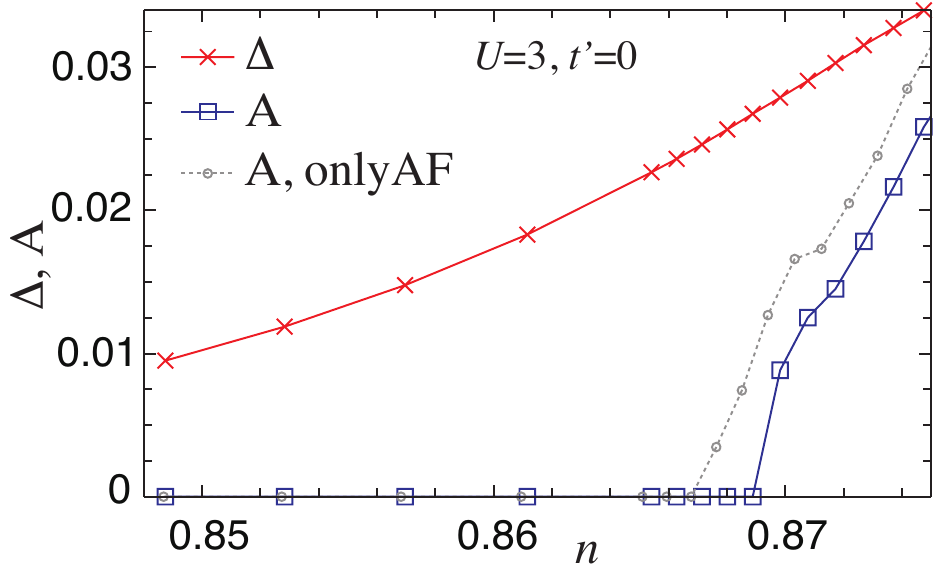}
\caption{Amplitudes of magnetic and superconducting gap functions in
 the ground state of the two-dimensional Hubbard model with $t'=0$ and
 $U=3$ in the density range $0.848 < n < 0.875$.
 The amplitude of the magnetic gap function obtained from a purely
 magnetic solution is also shown for comparison.}
\end{center}
\end{figure}

The extent of phase separated regions obviously depends sensitively on the model parameters. The issue of phase separation in the Hubbard model is discussed controversially in the literature.
Mean-field theory yields phase separation in the ground state at $t'=0$ near  half-filling for any $U > 0$, while a finite $t'$ is found to stabilize homogeneous solutions at weak interactions both in the electron- and hole-doped regime $[\rm S1]$.
This is in full agreement with our results.
At strong coupling, phase separation at $t'=0$ is found by a dual-fermion approach $[\rm S2]$, and on the hole-doped side at $t'=-0.3$ within a variational cluster approximation $[\rm S3]$.
By contrast, in a dynamical cluster calculation phase separation was found only for a positive $t'$ on the hole-doped side (equivalent to a negative $t'$ and electron-doping) $[\rm S4]$.
Quantum Monte Carlo calculations and exact diagonalization do not provide evidence for phase separation at all, but this may well be due to finite size effects.


\end{document}